\RequirePackage{fix-cm}

\documentclass[smallcondensed]{svjour3}     % onecolumn (ditto)
\smartqed  % flush right qed marks, e.g. at end of proof
\usepackage{graphicx}
\usepackage{enumerate}
\usepackage{CJK}
\usepackage{amsmath}
\usepackage{xcolor}
\usepackage{bbding}

\begin{document}

\title{Comment on ``Fault-Tolerate Quantum Private Comparison Based on GHZ States and ECC"}

\titlerunning{Comment on ``Fault-Tolerate Quantum Private Comparison Based on...}        % if too long for running head

\author{Sai Ji           \and
        Fang Wang        \and
        Wen-Jie Liu      \and
        Chao Liu         \and
        Hai-Bin Wang
}

\authorrunning{S. Ji, F. Wang, W.-J. Liu, etc.} % if too long for running head

\institute{S. Ji  \and W.-J. Liu \Envelope (Corresponding author) \at
              Jiangsu Engineering Center of Network Monitoring, Nanjing University of Information Science \& Technology, Nanjing 210044, China \\
              \email{wenjiel@163.com}
           \and
           S. Ji \and F. Wang  \and W.-J. Liu \and  C. Liu \and H.-B. Wang \at
              School of Computer and Software, Nanjing University of Information Science \& Technology, Nanjing 210044, China\\
}

\date{Received: 26 March 2014 / Accepted: date}

\maketitle

\begin{abstract}
A two-party quantum private comparison scheme using GHZ states and error-correcting code (ECC) was introduced in Li et al.'s paper [Int. J. Theor. Phys. 52: 2818-2815, 2013], which holds the capability of fault-tolerate and could be performed in a none-ideal scenario. However, this study points out there exists a fatal loophole under a special attack, namely the twice-Hadamard-CNOT attack. A malicious party may intercept the other's particles, firstly executes the Hadamard operations on these intercepted particles and his (her) own ones respectively, and then sequentially performs twice CNOT operations on them and the auxiliary particles prepared in advance. As a result, the secret input will be revealed without being detected through measuring the auxiliary particles. For resisting this special attack, an improvement is proposed by applying a permutation operator before TP sends the particle sequences to all the participants.
\end{abstract}

\section{Introduction}
\label{intro}
Since quantum mechanics principles are introduced into cryptography, quantum cryptography attracts more and more attention. Due to the characteristic of quantum unconditional security, many quantum cryptography protocols, such as quantum key distribution (QKD) [1-3], quantum direct communication (QDC) [4-7], quantum secret sharing (QSS) [8-10], quantum teleportation (QT) [11,12], have been proposed to solve various secure problems.

Recently, quantum private comparison (QPC) has become an important branch in quantum cryptography. Based on the properties of quantum mechanics, the participants can determine whether their secret inputs are equal or not without disclosing their own secrets to each other. In 2009, Yang et al. [13] put forward a pioneering QPC scheme based on Bell states and a hash function. Since then, a large number of QPC protocols utilizing the entangled states, such as EPR pairs, GHZ state, etc., have been proposed [14-23].

However, these QPC protocols [13-23] are feasible in the ideal scenario but not secure in practical scenario where faults (including noise and error) are existent in the quantum channel and measurement. In order to solve the problem, in 2013, Li et al. [24] present a novel QPC scheme based on GHZ states and error-correcting code (ECC) against noise. But, through analyzing Li et al.'s QPC scheme, we find it is unsecure under a special attack, called the twice-Hadamard-CNOT attack. To be specific, if any malicious party performed the twice-Hadamard-CNOT attack, he (she) can get another's secret input, which goes against the QPC's principles [25]. In order to fix the loophole, a simple solution by adopting a permutation operator before TP distributes the particles to the participants, is proposed.

The rest of this paper is constructed as follows. At first, Li et al.'s QPC protocol is briefly reviewed in Sect. 2. And in Sect. 3, we analyze the security of Li et al.'s QPC protocol by introducing the twice-Hadmard-CNOT attack, and give an improvement to fix the problem. Finally, a short collusion is drawn in Sect. 4.

\section{Review of Li et al.'s QPC protocol}
In the Ref. [24], in order to guarantee the QPC protocols secure in the practical scenario, Li et al. present a new two-party QPC scheme by using error-correcting code. The whole protocol consists of eight steps as below.

  \begin{enumerate}[(1)]
    \item
     Alice, Bob and Calvin prepare a [$m,n$] error-correcting code which uses $m$ bits codeword to encode $n$   bits word and can correct $l$ error bits in codeword with the error-correcting function $D(x^{m})$ according to the fault rate of the noise scenario. We suppose the error-correcting code's generator matrix is $G$, and check matrix is $Q$. Then they encode $X=(x_{0},x_{1},...,x_{n-1})$ and $Y=(y_{0},y_{1},...,y_{n-1})$ to $X'=(x'_{0},x'_{1},...,x'_{m-1})$ and $Y'=(y'_{0},y'_{1},...,y'_{m-1})$ with the generator matrix $G$, respectively. There are
     \begin{equation}
       X'=X\cdot G,
     \end{equation}
     \begin{equation}
        Y'=Y\cdot G.
     \end{equation}

     \item
     Calvin prepares $m$ triplet GHZ states all in
     \begin{equation}
       \begin{split}
       |\psi\rangle_{CAB}&=\frac{1}{\sqrt{2}}(|000\rangle+|111\rangle)_{CAB} \\
                         &=\frac{1}{2}(|+++\rangle+|+--\rangle+|-+-\rangle+|--+\rangle)_{CAB},
       \end{split}
     \end{equation}
where $|0\rangle$ and $|1\rangle$ are measured in $Z$ basis, $|+\rangle$ and $|-\rangle$ are measured in $X$ basis, and $|\pm\rangle=\frac{1}{\sqrt{2}}(|0\rangle\pm|1\rangle)$. Calvin divides these $m$  GHZ states into three sequences $S_{A}$, $S_{B}$ and $S_{C}$, which includes the first, the second, and the third particles of all GHZ states, respectively.
   \item
   Calvin prepares some decoy photons prepared in states ${|0\rangle,|1\rangle,|+\rangle,|-\rangle}$ in random. He inserts these decoy photons in $S_{A}$ and $S_{B}$ at random positions to form sequences $S^{*}_{A}$ and $S^{*}_{B}$ respectively. Calvin retains the quantum sequence $S_{C}$ and sends the sequence $S^{*}_{A}$ to Alice, $S^{*}_{B}$ to Bob.
   \item
   When Alice and Bob receive $S^{*}_{A}$ and $S^{*}_{B}$, Calvin announces the positions and measurement base of these decoy photons. Alice and Bob measure them in the same base and announce their outcome. If the error rate exceeds a rational threshold, Calvin aborts the protocol and restarts from Step (1). Otherwise, there is no eavesdropper, and the protocol continues to the next step.
   \item
   Alice and Bob recover $S_{A}$ and $S_{B}$ respectively by discarding the decoy photons. Then Alice, Bob and Calvin measure $S_{A}$, $S_{B}$ and $S_{C}$ in $X$ basis, respectively. If the measurement result is $|+\rangle$ ($|-\rangle$), then they encode it as the classical bit 0 (1). Thus, each of Alice, Bob and Calvin will obtain $m$ bits from $S_{A}$, $S_{B}$ and $S_{C}$, respectively. We denote each of them as $k^{A}_{i}$, $k^{B}_{i}$ and $k^{C}_{i}$ (i=0,1,...,m-1).
   \item
   Alice and Bob calculate $x^{''}_{i}=k^A_{i}\oplus x^{'}_{i}$ and $y^{''}_{i}=k^{B}_{i}\oplus y^{'}_{i}$. They announce $X^{''}=(x^{''}_{0},x^{''}_{1},...,x^{''}_{m-1})$ and $Y^{''}=(y^{''}_{0},y^{''}_{1},..,y^{''}_{m-1})$ to Calvin.
   \item
   Calvin calculates $c^{'}_{i}=k^C_{i}\oplus x^{''}_{i}\oplus y^{''}_{i}$, and gets $m$ bits sequence $C^{'}=(c^{'}_{0},c^{'}_{1},...,c^{'}_{m-1})$.
   \item
   Then Calvin uses the check matrix $Q$ to check whether the number of error bits exceeds the threshold $l$. If it does, Calvin aborts the protocol and restarts from Step (1). Otherwise, he gets $n$ bits sequence $C^{*}$ by decoding $C^{'}$  with error-correcting function $D(C^{'})$. If there is at least one bit 1 in $C^{*}$, Calvin announces $X\neq Y$. Otherwise, he announces $X=Y$.
  \end{enumerate}

  In Ref. [24], the authors claimed the scheme was secure even in the practical scenario. However, we will show how a malicious party gets the other's secret input by launching the twice-Hadmard-CNOT attack in the next section.

\section{Twice-Hadmard-CNOT attack on Li et al.'s QPC protocol and the improvement}
\subsection{Twice-Hadmard-CNOT attack on Li et al.'s QPC protocol}
As analyzed in Ref.[24], due to the decoy photons adopted in the Li et al.'s QPC protocol, some well-known attacks, such as intercept-resend attack, measurement-resend attack, and entanglement-resend attack, can be detected via the checking mechanism. Unfortunately, we find Li et al.'s QPC protocol cannot resist a special attack, i.e., the twice-Hadmard-CNOT attack. To be specific, if any party (Alice or Bob) performs the twice-Hadmard-CNOT attack, he/she can get the other's secret input. The detailed procedure of the twice-Hadmard-CNOT attack is depicted as follows.

Without loss of generality, we suppose Bob is malicious, who wants to get Alice's secret input. At first, Bob prepares $m$ auxiliary particles all in the state $|0\rangle_{e}$, then, the GHZ state and an auxiliary particle compose a composite system:
\begin{equation}
  \begin{split}
    |\eta\rangle&=|\psi\rangle_{CAB}|0\rangle_{e} \\
                &=\frac{1}{\sqrt{2}}(|0000\rangle+|1110\rangle)_{CABe} \\
                &=\frac{1}{2}(|+++0\rangle+|+--0\rangle+|-+-0\rangle+|--+0\rangle)_{CABe},
  \end{split}
\end{equation}
where the subscript $C$, $A$, $B$ represent the particles in the hand of Calvin, Alice, Bob, respectively, and the subscript $e$ represents the auxiliary particle.

In Step (3), when Calvin sends the sequence $S^{*}_{A}$ to Alice, Bob may intercept $S^{*}_{A}$ and execute a Hardmard ($H$) operation on every particle in $S^{*}_{A}$ to form sequence $S^{**}_{A}$. Then, he performs a controlled-NOT (CNOT) operation $C_{Ae}$ on every particle in $S^{**}_{A}$ and the corresponding auxiliary particle $e$. Here, the particle in $S^{**}_{A}$ is the control qubit, while particle $e$ is the target qubit. After that, Bob performs another $H$ operation on every particle in $S^{**}_{A}$ in order to restore sequence $S^{**}_{A}$ to $S^{*}_{A}$, and sends $S^{*}_{A}$ to Alice. What should be noted is that the transmitting sequence $S^{*}_{A}$ remains unchanged.

Since Calvin announces the decoy photons' positions in Step (4), Bob can discard the auxiliary particles $e$ which correspond to the decoy photons in $S^{*}_{A}$. And in Step (5), after Bob recovers $S_{B}$ by discarding the decoy photons, he executes an $H$ operation on every particle in $S_{B}$ to form sequence $S^{**}_{B}$. Then Bob performs a CNOT operation $C_{Be}$ on every particle (control particle) in $S^{**}_{B}$ and the corresponding auxiliary particle $e$ (target particle). After that, he performs a $H$ operation on every particle in $S^{**}_{B}$, which aims to restore $S^{**}_{B}$ to $S_{B}$, and every auxiliary particle, respectively. Now, the state of the composite system is changed into
\begin{equation}
  |\eta^{'}\rangle=\frac{1}{2}(|++\rangle_{Ce}(|++\rangle+|--\rangle)_{AB}+|--\rangle_{Ce}(|+-\rangle+|-+\rangle)_{AB}).
\end{equation}
From Eq.(5), a rule can be concluded: if $e$ is $|+\rangle$, particle $A$ and particle $B$ are in the same state; and if $e$ is $|-\rangle$, they are in the different states.

After Alice announces $X^{''}=(x^{''}_{0},x^{''}_{1},...,x^{''}_{m-1})$ in Step (6), Bob measures them in the $X$ basis, and get the measurement result $k^{e}_{i}(i=0,1,...,m-1)$. According to $k^{e}_{i}$ and the rule from Eq.(5), Bob can obtain Alice's states $k^{A}_{i}(i=0,1,...,m-1)$. Since $X^{'}$ has been announced by Alice, Bob can calculate $x^{''}_{i}\oplus k^{A}_{i}=(k^{A}_{i}\oplus x^{'}_{i})\oplus k^{A}_{i}=x^{'}_{i}$, that is to say, he can get Alice's secret input $x^{'}_{i}$.

For sake of clearness, the above procedure of the twice-Hadmard-CNOT attack can be intuitively demonstrated by the following figure (See Fig. 1).
\begin{figure}
  \includegraphics{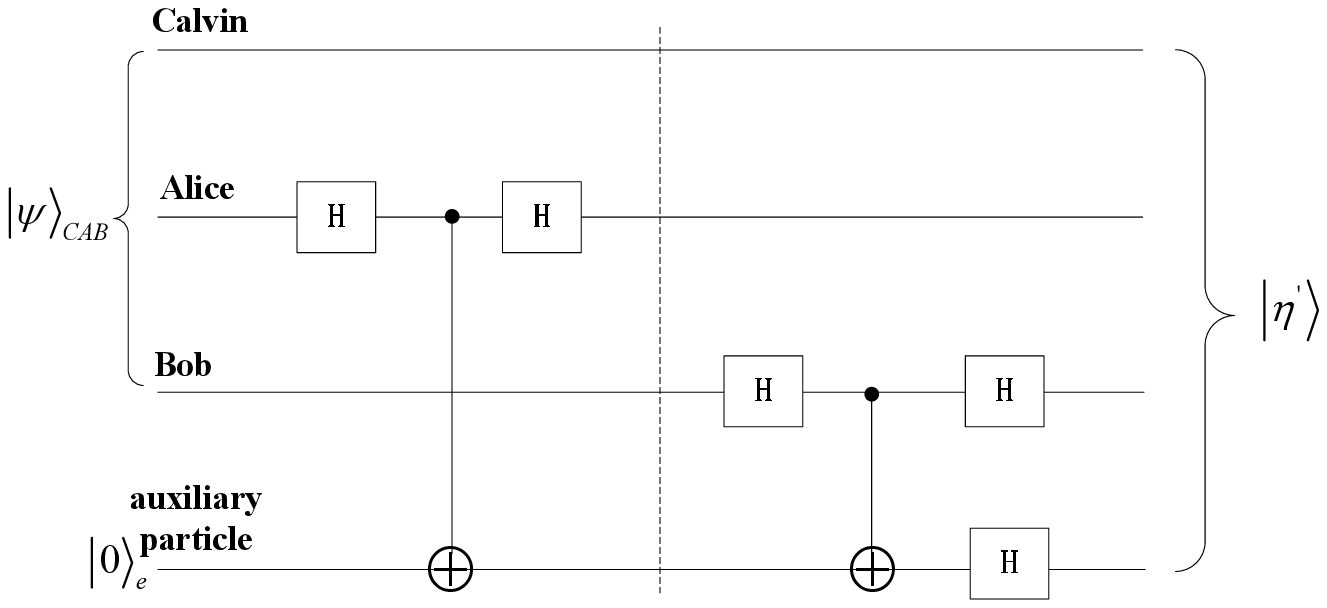}
\caption{The circuit diagram of the process of twice-Hadmard-CNOT attack. Here, $|\psi\rangle_{CAB}$ is a GHZ state in the state $(|000\rangle+|111\rangle)/\sqrt{2}$ shared by Calvin, Alice and Bob. $H$ is Hadmard operation, and \protect\includegraphics{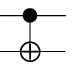} represents the controlled-NOT gate in which the top line denotes the control qubit, the bottom line the target qubit.}
\end{figure}

\subsection{The improvement}
In order to fix the loophole of Li et al.'s QPC protocol, we meliorate it by applying a permutation operator before TP sends the particle sequences to all the participants. To be specific, Step (3), Step (4) and Step (5) in the original protocol need to be revised as follows.
\begin{enumerate}
  \item [$(3)^{*}$]
  Calvin prepares two group $r$-length decoy photons sequences at random in $\{|+\rangle,|-\rangle,|0\rangle,|1\rangle\}$, namely $R_{A}$, $R_{B}$, and concatenates them with $S_{A}$ ($S_{B}$) to form an extended sequence $S^{'}_{A}=R_{A}||S_{A}$ ($S^{'}_{B}=R_{B}||S_{B}$), respectively. Then Calvin applies a permutation operator $\Pi_{(r+m)}$ on $S^{'}_{A}$ ($S^{'}_{B}$) to create a new sequence $\Pi_{(r+m)}S^{'}_{A}=S^{*}_{A}$ ($\Pi_{(r+m)}S^{'}_{B}=S^{*}_{B}$), and sends the new sequence $S^{*}_{A}$ to Alice, $S^{*}_{B}$ to Bob.

  \item[$(4)^{*}$]
  When Alice and Bob receive $S^{*}_{A}$ and $S^{*}_{B}$, Calvin announces the coordinates of the decoy qubits $\Pi_{r}$ ($\Pi_{r}\subset\Pi_{r+m}$) send by him, and the corresponding measurement bases. Note that Calvin does not disclose the actual order of the message qubits. Then Alice and Bob measure these decoy qubits in the same bases and announce their outcomes. If the error rate exceeds a rational threshold, Calvin aborts the protocol and restarts from Step (1); otherwise, there is no eavesdropper, and the protocol continues to the next step.

  \item[$(5)^{*}$]
  Alice and Bob discard their decoy photons, and denote the left qubits in their hands as $S^{''}_{A}$ and $S^{''}_{B}$ (i.e.,$S^{''}_{A}=\Pi_{m}S_{A},S^{''}_{B}=\Pi_{m}S_{B}$), respectively. Then Alice, Bob and Calvin measure $S^{''}_{A}$, $S^{''}_{B}$ and $S_{C}$ in the $X$ basis, respectively, and obtain $m$ bits, $k^{A}_{i}$, $k^{B}_{i}$ and $k^{C}_{i}$ (i=0,1,...,m-1 ), respectively. After they have completed measurement operation, Calvin immediately announces the actual order of the message qubits $\Pi_{m}$ ($\Pi_{m}\subset\Pi_{r+m}$). Using this information, Alice (Bob) can rearrange $k^{A}_{i}$ ($k^{B}_{i}$) in correspondence with the original order of $S_{A}$ ($S_{B}$).
\end{enumerate}

Let us examine the security of our improvement under the twice-Hadmard-CNOT attack. Similarly, we suppose the malicious Bob aims to get Alice's secret input. In Step $(3)^{*}$, Since Calvin disrupts the sequences $S^{'}_{A}$ and $S^{'}_{B}$ by using $\Pi_{(r+m)}$ to get $S^{*}_{A}$, $S^{*}_{B}$, it means the order of transmited sequences $S^{*}_{A}$, $S^{*}_{B}$ is fully disturbed. And after discarding the decoy qubits $R_{A}$ and $R_{B}$ in Step $(5)^{*}$, Alice and Bob can only get $S^{''}_{A}$ and $S^{''}_{B}$, and cannot recover the original sequences $S_{A}$ and $S_{B}$ because that $\Pi_{m}$ is only known by Calvin. So, even if Bob launch the twice-Hadmard-CNOT attack, he cannot get the final state $|\eta^{'}\rangle=\frac{1}{2}(|++\rangle_{Ce}(|++\rangle+|--\rangle)_{AB}+|--\rangle_{Ce}(+-\rangle+|-+\rangle)_{AB})$. That is to say, Bob cannot get the correlations between Alice's and Bob's qubits by measuring the auxiliary particles $e$.

For simplicity, we take a simple two-bit private comparisonas example without considering error-correcting code and the decoy photos. In the case, suppose Alice's input is 10, Bob's input is 11, and Calvin prepares two GHZ states all in the state as Eq.(3),
\begin{equation}
\begin{split}
  |\psi\rangle_{C_{1}A_{1}B_{1}}\otimes|\psi\rangle_{C_{2}A_{2}B_{2}}
  &=1/2(|+++\rangle+|+--\rangle+|-+-\rangle+|--+\rangle)_{C_{1}A_{1}B_{1}}\\
  &\otimes1/2(|+++\rangle+|+--\rangle+|-+-\rangle+|--+\rangle)_{C_{2}A_{2}B_{2}}
\end{split}
\end{equation}
In Step $(3)^{*}$, Calvin executes a permutation operator $\Pi_{m=2}$ on sequence $S^{'}_{A}$ and $S^{'}_{B}$, then the system may be changed into
\begin{equation}
\begin{split}
 \Pi_{m=2}(|\psi\rangle_{C_{1}A_{1}B_{1}}\otimes|\psi\rangle_{C_{2}A_{2}B_{2}})&=|\psi\rangle_{C_{1}A_{2}B_{2}}\otimes|\psi\rangle_{C_{2}A_{1}B_{1}}\\
 &=1/4\{|+\rangle_{C_{1}}(|++\rangle+|--\rangle)_{A_{2}B_{2}}\\
 &\quad\;\otimes|+\rangle_{C_{2}}(|++\rangle+|--\rangle)_{A_{1}B_{1}}\\
 &\quad\;+|+\rangle_{C_{1}}(|+-\rangle+|-+\rangle)_{A_{2}B_{2}}\\
 &\quad\;\otimes|-\rangle_{C_{2}}(|++\rangle+|--\rangle)_{A_{1}B_{1}}\\
 &\quad\;+|-\rangle_{C_{1}}(|++\rangle+|--\rangle)_{A_{2}B_{2}}\\
 &\quad\;\otimes|+\rangle_{C_{2}}(|+-\rangle+|-+\rangle)_{A_{1}B_{1}}\\
 &\quad\;+|-\rangle_{C_{1}}(|+-\rangle+|-+\rangle)_{A_{2}B_{2}}\\
 &\quad\;\otimes|-\rangle_{C_{2}}(+-\rangle+|-+\rangle)_{A_{1}B_{1}}
 \}
\end{split}
\end{equation}
After Bob performs the twice-Hadmard-CNOT attack, the composite system which consists of GHZ state and auxiliary particle becomes
\begin{equation}
  \begin{split}
    |\eta^{'}\rangle_{1}\otimes|\eta^{'}\rangle_{2}
    &=1/4\{|+\rangle_{C_{1}}|+\rangle_{e_{1}}(|++\rangle+|--\rangle)_{A_{2}B_{2}}\\
    &\quad\;\otimes|+\rangle_{C_{2}}|+\rangle_{e_{2}}(|++\rangle+|--\rangle)_{A_{1}B_{1}}\\ &\quad\;+|+\rangle_{C_{1}}|-\rangle_{e_{1}}(|+-\rangle|-+\rangle)_{A_{2}B_{2}}\\
    &\quad\;\otimes|-\rangle_{C_{2}}|+\rangle_{e_{2}}(|++\rangle+|--\rangle)_{A_{1}B_{1}}\\
    &\quad\;+|-\rangle_{C_{1}}|+\rangle_{e_{1}}(|++\rangle+|--\rangle)_{A_{2}B_{2}}\\
    &\quad\;\otimes|+\rangle_{C_{2}}|-\rangle_{e_{2}}(|+-\rangle+|-+\rangle)_{A_{1}B_{1}}\\
    &\quad\;+|-\rangle_{C_{1}}|-\rangle_{e_{1}}(|+-\rangle+|-+\rangle)_{A_{2}B_{2}}\\
    &\quad\;\otimes|-\rangle_{C_{2}}|-\rangle_{e_{2}}(|+-\rangle+|-+\rangle)_{A_{1}B_{1}}
    \}.
  \end{split}
\end{equation}
From above equation, we cannot get the correlations between the states of $\{A_{1},B_{1}\}$ or $\{A_{2},B_{2}\}$ according to the the final state of the auxiliary particle $e_{1}$ or $e_{2}$. That means, Bob cannot steal Alice's input through measuring the auxiliary particles. So, we can say the improvement can resist the twice-Hadmard-CNOT attack.

\section{Conclusione}
In all the QPC protocols, in order to ensure the protocol's security, we must guarantee any participant only knows his (her) own secret input without obtaining another' secret input. In this paper, we firstly review and analyze Li et al.'s two-party QPC protocol, and find it cannot resist the twice-Hadmard-CNOT attack, i.e., if one participant Bob launches this attack, he (she) can get the other's secret input without being detected. For avoiding this loophole, we adopt the permutation operator to rearranges the quantum sequences sent to Alice and Bob from Charlie. The delicate analysis shows the security of our improvement can be guaranteed well and truly.\\\\
\textbf{Acknowledgments}\quad\;\small\textmd{This work is supported by the National Nature Science Foundation of China (Grant Nos. 61103235, 61373131 and 61373016), the Priority Academic Program Development of Jiangsu Higher Education Institutions (PAPD), and State Key Laboratory of Software Engineering, Wuhan University (SKLSE2012-09-41).}


\begin{thebibliography}{}
\bibitem{RefB}
Bennett, C.H., Brassard, G.: Quantum cryptography: public-key distribution and coin tossing. In: Proceedings of IEEE International conference on Computers, Systems and Signal Processing Bangalore, New York, pp. 175-179. Bangalore, India, (1984)
\bibitem{RefJ}
Long, G.L., Liu, X.S.: Theoretically efficient high-capacity quantum-key-distribution scheme. Phys. Rev. A. \textbf{65}, 032302 (2002). doi:\textcolor{blue}{10.1103/PhysRevA.65.032302}
\bibitem{RefJ}
Moroder, T., Curty, M., Lim, C.C.W., Thinh, L.P., Zbinden, H., Gisin, N.: Security of Distributed-Phase-Reference Quantum Key Distribution. Phys. Rev. Lett. \textbf{109}, 260501 (2012)
\bibitem{RefJ}
Deng, F.G., Long, G.L., Liu, X.S.: Two-step quantum direct communication protocol using the Einstein-Podolsky-Rosen pair block. Phys. Rev. A. \textbf{68}, 042317 (2003). doi:\textcolor{blue}{10.1103/PhysRevA.68.042317}
\bibitem{RefJ}
Deng, F.G., Long, G.L.: Secure direct communication with a quantum one-time pad. Phys. Rev. A. \textbf{69}, 052319 (2004). doi:\textcolor{blue}{10.1103/PhysRevA.69.052319}
\bibitem{RefJ}
Liu, W.J., Chen, H.W., Li, Z.Q., Liu, Z.H.: Efficient quantum secure direct communication with authentication. Chinese Physics Letters. \textbf{25}, 2354-2357 (2008).
\bibitem{RefJ}
Liu, W.J., Chen, H.W., Ma, T.H., Li, Z.Q., Liu, Z.H., Hu, W.B.: An efficient deterministic secure quantum communication scheme based on cluster states and identity authentication. Chin. Phys. B. \textbf{18}, 4105-4109 (2009). doi:\textcolor{blue}{10.1088/1674-1056/18/10/007}
\bibitem{RefJ}
Cleve, R., Gottesman, D., Lo, H.K.: How to share a quantum secret. Phys. Rev. Lett. \textbf{83}, 648-651 (1999). doi:\textcolor{blue}{10.1103/PhysRevLett.83.648}
\bibitem{RefJ}
Xu, J., Chen, H.W., Liu, W.J., Liu, Z.H.: Selection of unitary operations in quantum secret sharing without entanglement. Sci. China Inf. Sci. \textbf{54}, 1837-1842 (2011). doi:\textcolor{blue}{10.1007/s11432-011-4240-9}
\bibitem{RefJ}
Wang, H.B., Huang, Y.G., Fang, X., Gu, B., Fu, D.S.: High-Capacity Three-Party Quantum Secret Sharing with Single Photons in Both the Polarization and the Spatial-Mode Degrees of Freedom. Int. J. Theor. Phys. \textbf{52}, 1043-1051 (2013). doi:\textcolor{blue}{10.1007/s10773-012-1418-x}
\bibitem{RefJ}
Bouwmeester, D., Pan, J.W., Mattle, K., Eibl, M., Weinfurter, H., Zeilinger, A.: Experimental quantum teleportation. Nature \textbf{390}, 575-579 (1997).
\bibitem{RefJ}
Furusawa, A., Sorensen, J.L., Braunstein, S.L., Fuchs, C.A., Kimble, H.J., Polzik, E.S.: Unconditional quantum teleportation. Science \textbf{282}, 706-709 (1998). doi:\textcolor{blue}{10.1126/science.282.5389.706}
\bibitem{RefJ}
Yang, Y.G., Wen, Q.Y.: An efficient two-party quantum private comparison protocol with decoy photons and two-photon entanglement. J. Phys. A: Math. Theor. \textbf{42}, 055305 (2009).doi:\textcolor{blue}{10.1088/1751-8113/42/5/055305}
\bibitem{RefJ}
Chen, X.B., Xu, G., Niu, X.X., Wen, Q.Y., Yang, Y.X.: An efficient protocol for the private comparison of equal information based on the triplet entangled state and single-particle measurement. Opt. Commun. \textbf{283}, 1561-1565 (2010). doi:\textcolor{blue}{10.1016/j.optcom.2009.11.085}
\bibitem{RefJ}
Jia, H.Y., Wen, Q.Y., Li, Y.B., Gao, F.: Quantum Private Comparison Using Genuine Four-Particle Entangled States. Int. J. Theor. Phys. \textbf{51}, 1187-1194 (2012). doi:\textcolor{blue}{10.1007/s10773-011-0994-5}
\bibitem{RefJ}
Liu, W., Wang, Y.B.: Quantum Private Comparison Based on GHZ Entangled States. Int. J. Theor. Phys. \textbf{51}, 3596-3604 (2012). doi:\textcolor{blue}{10.1007/s10773-012-1246-z}
\bibitem{RefJ}
Liu, W., Wang, Y.B., Cui, W.: Quantum Private Comparison Protocol Based on Bell Entangled States. Commun. Theor. Phys. \textbf{57}, 583-588 (2012). doi:\textcolor{blue}{10.1088/0253-6102/57/4/11}
\bibitem{RefJ}
Tseng, H.Y., Lin, J., Hwang, T.: New quantum private comparison protocol using EPR pairs. Quantum Inf. Process. \textbf{11}, 373-384 (2012). doi:\textcolor{blue}{10.1007/s11128-011-0251-0}
\bibitem{RefJ}
Sun, Z.W., Long, D.Y.: Quantum Private Comparison Protocol Based on Cluster States. Int. J. Theor. Phys. \textbf{52}, 212-218 (2013). doi:\textcolor{blue}{10.1007/s10773-012-1321-5}
\bibitem{RefJ}
Lin, J., Yang, C.-W., Hwang, T.: Quantum private comparison of equality protocol without a third party. Quantum Inf. Process. \textbf{13}, 239-247 (2014). doi:\textcolor{blue}{10.1007/s11128-013-0645-2}
\bibitem{RefJ}
Liu, W.J., Liu, C., Liu, Z.H., Liu, J.F., Geng, H.T.: Same Initial States Attack in Yang et al.'s Quantum Private Comparison Protocol and the Improvement. Int. J. Theor. Phys. \textbf{53}, 271-276 (2014). doi:\textcolor{blue}{10.1007/s10773-013-1807-9}
\bibitem{RefJ}
Liu, W.J., Liu, C., Chen, H.W., Liu, Z.H., Yuan, M.X., Lu, J.S.: IMPROVEMENT ON ``AN EFFICIENT PROTOCOL FOR THE QUANTUM PRIVATE COMPARISON OF EQUALITY WITH W STATE". Int. J. Quantum Inf. \textbf{0}, 1450001 (2014). doi:\textcolor{blue}{10.1142/S0219749914500014}
\bibitem{RefJ}
Liu, W.J., Liu, C., Wang, H.b., Liu, J.F., Wang, F., Yuan, X.M.: Secure Quantum Private Comparison of Equality Based on Asymmetric W State. Int. J. Theor. Phys., 1-10 (2014). doi:\textcolor{blue}{10.1007/s10773-013-1979-3}
\bibitem{RefJ}
Li, Y.B., Wang, T.Y., Chen, H.Y., Li, M.D., Yang, Y.T.: Fault-Tolerate Quantum Private Comparison Based on GHZ States and ECC. Int. J. Theor. Phys. \textbf{52}, 2818-2825 (2013). doi:\textcolor{blue}{10.1007/s10773-013-1573-8}
\bibitem{RefJ}
Liu, W.J., Liu, C., Wang, H.B., Jia, T.T.: Quantum Private Comparison: A Review. IETE Tech. Rev. \textbf{30}, 439-445 (2013). doi:\textcolor{blue}{10.4103/0256-4602.123129}

\end{thebibliography}
\end{document}